\documentclass[aps,prl,twocolumn,groupedaddress,showpacs]{revtex4}
\usepackage{graphicx}
\usepackage{wasysym}
\usepackage[dvips]{color}

\begin{document}
\title{The Boson-Hubbard Model on a Kagome Lattice with Sextic Ring-Exchange Terms}

\author{Val\'ery G.~Rousseau}
\author{Ka-Ming Tam}
\author{Mark Jarrell}
\author{Juana Moreno}
\affiliation{Department of Physics \& Astronomy and Center for Computation \& Technology, 
Louisiana State University, Baton Rouge, Louisiana 70803, USA}

\relpenalty=10000      
\binoppenalty=10000


\begin{abstract}
High order ring-exchange interactions are crucial for the study of quantum fluctuations on many highly frustrated systems. 
A versatile and efficient quantum Monte Carlo method, which can handle
finite and essentially zero temperature and canonical and grand-canonical ensembles, has
long been sought. In this paper, we present the first exact quantum Monte Carlo study of a model of hard-core bosons with sixth 
order ring-exchange interactions on a two-dimensional kagome lattice. 
By using the \textit{Stochastic Green Function} algorithm with \textit{global space-time update},
we show that the system becomes unstable in the limit of large ring-exchange interactions. It undergoes a 
phase separation at all fillings, except at $\frac13$ and $\frac23$ fillings for which the superfluid density vanishes 
and an unusual mixed valence bond and charge density ordered solid is formed. This explains the universal features seen
in previous studies on various different models, such as the transverse field Ising models, on a kagome 
lattice near the classical limit.
\end{abstract}

\maketitle

Interest in ring-exchange interactions in quantum many-body systems has a long history originating from the study 
of quantum solids, a typical example being solid Helium-3 ~\cite{Thouless,Herring}. Recently the study of 
ring-exchange interactions has resurged with boson and spin models~\cite{Balents,Paramekanti,Sandvik,Rousseau1,Melko,Rousseau2}. 
In particular, multiple particle exchange has been suggested as a possible candidate to induce a normal ``Bose 
metal" or ``Bose liquid" phase  at zero temperature, in which there are no broken symmetries associated with 
superfluidity or charge density wave phases~\cite{Paramekanti,Hermele,Nikolic,Nikolic-2005}. 
Studies on a square lattice with four-site ring-exchange plaquettes suggest that ordered phases always 
dominate~\cite{Sandvik,Melko,Rousseau1,Rousseau2}. However, these results still leave the possibility that a 
Bose liquid or spin liquid phase may exist in certain frustrated lattices~\cite{Balents,Dang}.

Here we explore a frustrated kagome lattice built from corner sharing triangles in two dimensions. 
The most important characteristic of highly frustrated quantum systems is their ground state degeneracy in the 
classical limit~\cite{Villain-1979,Diep}. In such systems the classical ground state manifold, which is given by the least 
frustrated Ising spin configurations, results in local constraints in each unit. One of the prominent problems in frustrated 
quantum magnetism is whether quantum fluctuations partially alleviate the classical ground state degeneracy via the 
order-by-disorder mechanism or by forming a quantum spin liquid driven by quantum fluctuations~\cite{Villain-1980,Shender,Diep}. 
A natural route to study this problem is to construct an effective theory for the low energy quantum fluctuations which is 
confined to the Hilbert space of the degenerate classical ground state manifold. From the strong coupling perturbation theory, 
the leading term for the quantum fluctuations involves multiple-spin loop flips which form the ring-exchange term.  
For kagome and pyrochlore lattices, the smallest ring-exchange term appears at the sixth order~\cite{Nikolic-2005,Hermele,Shannon}. 
Therefore, the study of quantum fluctuations on some of the most important highly frustrated lattices naturally involves 
effective models with multiple-spin ring-exchange.

Recently those ring-exchange models have been formulated in terms of gauge theories \cite{Hermele,Nikolic,Neto}. 
The gauge theory formulation is essentially the manifestation of the constraint for the projected Hilbert space within 
the classical degenerate manifold, and the effective gauge theory is, in turn, studied by a duality mapping. Various
phases have been proposed for different models based on this type of calculations. 

It is extremely important 
to have a systematic unbiased numerical method to test the various proposals for exotic ordered valence 
bond and disordered spin liquid phases. However, numerical studies on those models have proved to be rather 
difficult. There are some recent studies using the World-line and the Stochastic Series Expansion algorithms 
with loop update schemes on four-site ring-exchange models~\cite{Rousseau1,Sandvik,Melko,Rousseau2}. For these models, 
ring-exchange interactions are described by a term that performs a correlated hopping of two particles in opposite 
directions, with no contribution to the winding. Thus one can expect the superfluidity to be destroyed when this term 
is dominant. However, instead of the proposed Bose liquid phase, these exact studies have found charge density waves, 
valence bond solids, or phase separation.

\begin{figure}[h]
  \centerline{\includegraphics[width=0.45\textwidth]{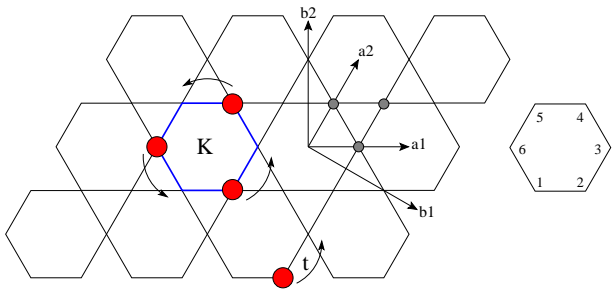}}
  \caption{(Color online) The kagome lattice and the effect of the different terms in the Hamiltonian. Each site 
  is shared by two hexagons, so the total number of sites is three times the number of hexagons (for a periodic lattice). 
  The usual kinetic term $t$ allows the particles to hop between near-neighboring sites. The ring-exchange term $K$ 
  performs a correlated hopping of three particles within the same hexagon. This process is possible only if the hexagon 
  contains exactly three non-near-neighboring particles.  The figure also shows our convention for the labels of the 
  sites of a given hexagon.}
  \label{KagomeLattice}
\end{figure}

The model we consider in this work consists of hard-core bosons on a two-dimensional kagome lattice 
(Fig.~\ref{KagomeLattice}). The underlying Bravais lattice is spanned by the basis vectors $(\vec a_1,\vec a_2)$ 
with lengths chosen as unity, and the kagome lattice is obtained by duplicating a set of three sites 
$S_1(\frac12,0)$, $S_2(0,\frac12)$, $S_3(\frac12,\frac12)$ (gray symbols). The reciprocal lattice is spanned by 
the vectors $(\vec b_1,\vec b_2)$, each with length $4\pi/\sqrt3$.  

The Hamiltonian takes the form (we use periodic boundary conditions)
\begin{eqnarray}
  \nonumber \hat\mathcal H= &-& t\sum_{\langle i,j\rangle}\big(a_i^\dagger a_j^{\phantom\dagger}+H.c.\big)\\
  \label{Hamiltonian}       &-& K\sum_{\hexagon}\big(a_1^\dagger a_3^\dagger a_5^\dagger a_6^{\phantom\dagger} a_4^{\phantom\dagger} a_2^{\phantom\dagger}+H.c.\big),
\end{eqnarray}
where $a_i^\dagger$ and $a_i^{\phantom\dagger}$ are the creation and annihilation operators of a hard-core 
boson on site $i$. These operators satisfy fermionic anti-commutation rules when acting on the same site, 
$a_i^2=0$, $a_i^{\dagger 2}=0$, $\big\lbrace a_i^{\phantom\dagger},a_i^\dagger\big\rbrace=1$, and bosonic 
commutation rules when acting on different sites, $\big[a_i^{\phantom\dagger},a_j^{\phantom\dagger}\big]=0$, $\big[a_i^{\dagger},a_j^{\dagger}\big]=0$,
$\big[a_i^{\phantom\dagger},a_j^{\dagger}\big]=0$. 
The usual kinetic term allows the particles to hop between near-neighboring sites $i$ and $j$. A sextic 
ring-exchange term allows three particles to perform a correlated hopping within the same hexagon, with
the $\sum_{\hexagon}$ over all hexagons. We study the model as a function of filling factor and $K/t$.

This model can be mapped to the $U(1)$ lattice gauge theory with a softened hard-core constraint, which 
plays a crucial role in the study of quantum fluctuations in some highly frustrated models~\cite{Hermele,Nikolic-2005}. 
Notably, similar models can possibly be realized in optical lattices with dipolar bosons~\cite{Tewari}. 
Since the sextic ring-exchange term couples six different sites at a time, an 
analytical treatment of the Hamiltonian (\ref{Hamiltonian}) is rather complicated~\cite{Nikolic-2005,Hermele}.
On the other hand, most current quantum Monte Carlo (QMC) methods used on a 
square lattice~\cite{Rousseau1,Sandvik,Melko,Rousseau2} need to decompose the Hamiltonian as a sum of two-site
coupling terms, which is not possible in our case. These methods need such a decomposition because they
use a {\it loop update} that consists of building a closed path in space and imaginary time, and 
raising or lowering the occupation number of the sites that are visited. By construction, these loops can 
update only one or two sites at a time, and are therefore suitable only for one-site or two-site coupling terms. 
In brief, the treatment of fourth order ring-exchange terms in previous studies required some special 
developments~\cite{Sandvik,Rousseau2} that are not easy to generalize to sixth or higher order. 

However, the {\it Stochastic Green Function} (SGF) algorithm~\cite{SGF,DirectedUpdate} with global space-time update~\cite{GlobalSpaceTimeUpdate} 
does not make use of such loop updates, instead, it performs a direct sampling of the partition function
by distributing Hamiltonian terms randomly in space and imaginary time. No particular decomposition of the 
Hamiltonian is required. Therefore the update procedure is totally independent of the structure of the 
Hamiltonian, and can be applied to any $n$-site coupling term with efficiency. In this work, we use the 
SGF algorithm and perform simulations of systems with sizes up to $18\times18$ hexagons (972 sites).

While the SGF algorithm is designed to work in the canonical ensemble (CE), an extension~\cite{GlobalSpaceTimeUpdate} 
allows us to simulate the grand-canonical ensemble (GCE). In the following we take advantage of this 
flexibility. We find convenient to use the GCE for studying the stability of the system. However,
it is much easier to use the CE on parameter regions where the system is stable only 
at specific densities. This is, to the best of our knowledge, the first time that an unbiased QMC method 
allows the simulation of sixth order coupling terms in both the CE and GCE at both finite and essentially 
zero temperature. For the GCE we add the usual term $-\mu\hat\mathcal N$ to the Hamiltonian~(\ref{Hamiltonian}), 
where $\mu$ is the chemical potential and $\hat\mathcal N$ is the total particle number operator. In the CE 
$\mu$ is not a control parameter, but it is measured at zero temperature as $\mu(N)=E(N+1)-E(N)$, where 
$E=\big\langle\hat\mathcal H\big\rangle$ and $N$ is the controlled number of particles.

In the following, we show that the model contains valence bond solids at densities $\rho=\frac13$ 
and $\rho=\frac23$ when the ring-exchange term is dominant. For our study, we will consider the superfluid density
\begin{equation}
  \rho_s=\frac{\big\langle\hat W_1^2+\hat W_2^2\big\rangle}{4t\beta},
\end{equation}
where $\beta$ is the inverse temperature and $\hat W_1$ and $\hat W_2$ are the winding numbers measured 
in the two directions $\vec a_1$ and $\vec a_2$. We also consider the structure factor 
$S(\vec k)=\big\langle\big|\tilde n(\vec k)\big|^2\big\rangle$, with
\begin{equation}
  \label{FourierTransform} \tilde n(\vec k)=\frac{1}{3L^2}\sum_p\big(\hat n_p-\rho\big)e^{-i\vec k\cdot\vec r_p},
\end{equation}
where $L$ is the linear system size (counting hexagons), $\rho$ is the density, 
$\hat n_p=a_p^\dagger a_p^{\phantom\dagger}$ is the number operator on site $p$, and $\vec r_p$ is
the position of site $p$. Note that the subtraction of $\rho$ in~(\ref{FourierTransform}) is meant 
to get rid of the lattice Bragg peaks.  In order to study the ground state properties, we systematically 
use the inverse temperature $\beta=2L/t$.

\begin{figure}[h]
  \centerline{\includegraphics[width=0.45\textwidth]{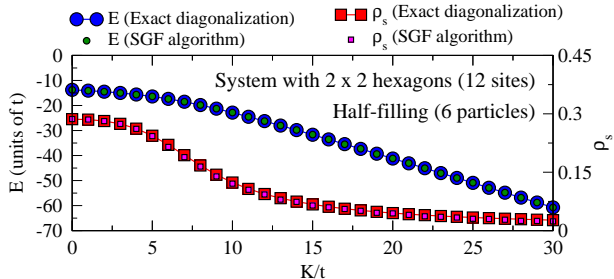}}
  \caption{(Color online) Comparison between an exact diagonalization and the SGF algorithm for the energy 
  $E$ and the superfluid density $\rho_s$.}
  \label{ExactDiagonalization}
\end{figure}

We start by illustrating in Fig.~\ref{ExactDiagonalization} the exactness of the SGF results, by comparing 
them with an exact diagonalization on a lattice with $2\times 2$ hexagons (12 sites) and six particles. The 
figure shows an excellent agreement for the energy $E$ and the superfluid density $\rho_s$.

\begin{figure}[h]
  \centerline{\includegraphics[width=0.45\textwidth]{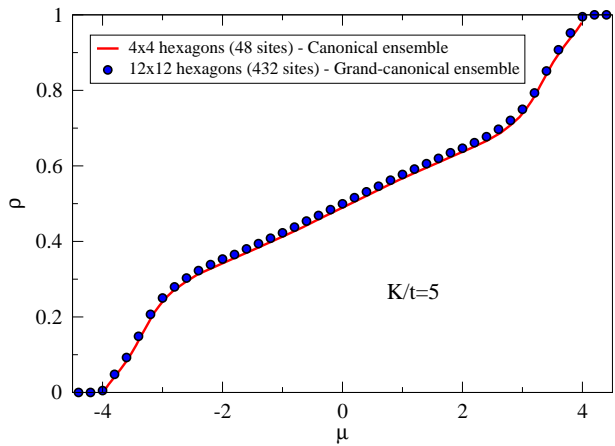}}
  \caption{(Color online) The total density $\rho$ as a function of the chemical potential $\mu$ for 
  $K=5t$. Results for a lattice with $4\times4$ hexagons in the canonical ensemble and $12\times12$ 
  hexagons in the grand-canonical ensemble are shown. The agreement between the two curves reveals 
  that finite size effects are sufficiently small to ensure the equivalence of the two ensembles.}
  \label{RhoVsMuK05}
\end{figure}

It is useful to analyze first the stability of the system. This is easily done by looking at the total 
density $\rho$ as a function of the chemical potential $\mu$. The slope of this function is proportional 
to the compressibility of the system, so an instability of the system results either in a negative slope 
in the CE, or a discontinuity of the total density in the GCE. We first discuss the small $K$ region, and
search for any phase transition as we increase the value of $K$. 
Figure~\ref{RhoVsMuK05} shows results for $K=5t$ and a system with $4\times4$ hexagons (48 sites) in the 
CE and $12\times12$ hexagons (432 sites) in the GCE. The agreement 
between the two curves reveals that finite size effects are sufficiently small to ensure the equivalence 
of the two ensembles.  For this strength of the ring-exchange interactions the slope of the curve 
remains positive and finite at all fillings, so there is no instability.

\begin{figure}[h]
  \centerline{\includegraphics[width=0.45\textwidth]{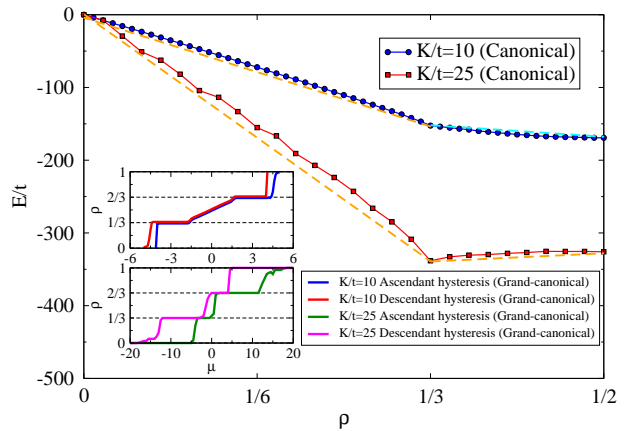}}
  \caption{(Color online) The energy as a function of the density for $K/t=10$ and $K/t=25$ in the CE. The insets show for comparison
  the density as a function of the chemical potential in the GCE, with a numerical hysteresis. See text for details.}
  \label{RhoVsMuK10K25}
\end{figure}

The situation changes as $K$ increases. Fig.~\ref{RhoVsMuK10K25} shows the total energy $E$ as a function of the density $\rho$
in the CE for $K/t=10$ and $K/t=25$. We show only data for $\rho\leq\frac12$, since the data for $\rho>\frac12$ can be deducted
by particle-hole symmetry. The cyan dotted line serves as a guide to indicate that the curvature of the
$K/t=10$ curve is positive for $\rho\in\big[\frac13;\frac23\big]$, while the orange dotted lines emphasize regions with
negative curvatures. Systems for which the energy has a negative second derivative with respect to the
density are thermodynamically unstable and undergo a phase separation. We also note the presence of a kink at $\rho=\frac13$ for
both curves, which indicates a gapped phase. We conclude that for $K/t=10$ the system is compressible for $\rho\in\big]\frac13;\frac23\big[$
and in a solid phase for $\rho=\frac13$ and $\rho=\frac23$, while it is unstable for all other densities. For $K/t=25$ the system
is unstable for all densities, except for $\rho=\frac13$ and $\rho=\frac23$ for which the phase is solid. The insets correspond
to GCE results that are in agreement and which, in addition to showing ``jumps" over the ``forbidden" densities, show the existence of a
``numerical hysteresis", depending
on if the initial state is empty (ascendant hysteresis) or full (descendant hysteresis). This phenomenon is observed in many
Monte Carlo studies~\cite{Beach,Hebert}, including the four-site ring-exchange model~\cite{Melko}. For these
parameters the grand-canonical algorithm is no longer able to sample all contributing states, which is an indication that the
system undergoes a spontaneous symmetry breaking in the thermodynamic limit. This also supports the importance of the ability
of the SGF algorithm to work in the canonical ensemble.

\begin{figure}[h]
  \centerline{\includegraphics[width=0.45\textwidth]{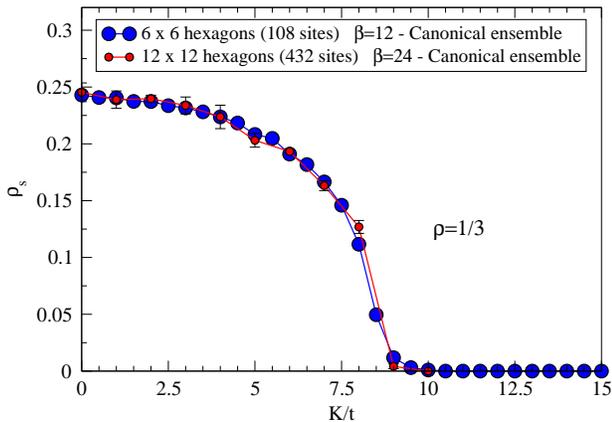}}
  \caption{(Color online) The superfluid density $\rho_s$ as a function of $K/t$ for $\rho=\frac13$. The 
   superfluid density is completely destroyed at the critical value $K_c\approx 9t$.}
  \label{RhosVsK}
\end{figure}

Since only the fillings $\rho=\frac13$ and $\rho=\frac23$ are stable in the large $K$ limit, it is 
convenient to analyze how the superfluid density $\rho_s$ is destroyed as a function of $K/t$ by working 
in the CE. We show in Fig.~\ref{RhosVsK} results for $\rho=\frac13$ only, the case with $\rho=\frac23$ being 
identical because of the particle-hole symmetry. There exists a critical value of the ring-exchange interaction, 
$K_c\approx 9t$ where the superfluid density vanishes. We show results for $6\times6$ hexagons (108 sites)
and $12\times12$ hexagons (432 sites) to illustrate that finite size effects are small.

\begin{figure}[htb]
  \centerline{\includegraphics[width=0.45\textwidth]{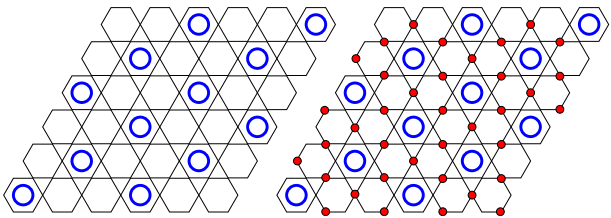}}
  \caption{(Color online) Two possible ordering patterns for $\rho=\frac13$ (left) and $\rho=\frac23$ 
  (right). The circles inside the hexagons represent three resonating particles.  The red dots represent 
  localized particles. We note that both orderings have threefold degeneracy. These two uniform ordering 
  patterns are only formed for $\rho=\frac13$ and $\rho=\frac23$. Other fillings are unstable towards
  phase separation.}
  \label{VBS}
\end{figure}

For $t \gg K$, the model is in the free hard-core limit, thus the superfluid phase prevails against solid or 
valence bond ordering. In the opposite $t \ll K$ limit, we expect the ground state to be formed by 
configurations that maximize the number of resonating hexagons.  Because a resonance can occur only if 
the hexagon contains exactly three non-first-neighboring particles, it is not possible to have two resonating
neighboring hexagons at first order in $K$.  This suggests that a phase with a maximum number of resonating 
hexagons can be formed at $\rho=\frac13$ by having all resonating hexagons surrounded by empty hexagons 
(Fig.~\ref{VBS} left), or at $\rho=\frac23$ by having all resonating hexagons surrounded by hexagons that 
contain three localized particles on the vertices of the triangles that are not shared with the resonating 
hexagons. (Fig.~\ref{VBS} right). These phases can be understood as a form of valence bond solid in which 
each valence bond now involves six sites, in contrast with the usual two-site singlets. We note that the phase
has been suggested in Kagome lattice on other models, such as the transverse field Ising model~\cite{Nikolic-2005,Blote}.

\begin{figure}[h]
  \centerline{\includegraphics[width=0.45\textwidth]{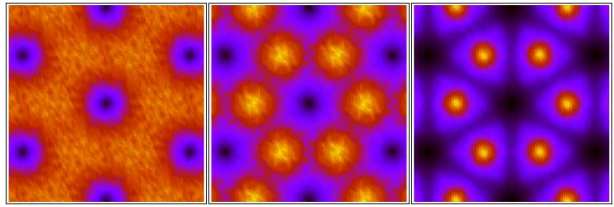}}
  \caption{(Color online) The structure factor $S(\vec k)$ for $18\times18$ hexagons (972 sites), 
  $\rho=\frac13$ and $K=0$ (left panel), $K=5t$ (middle panel), and $K=15t$ (right panel). The black 
  regions correspond to the locations of the lattice Bragg peaks. As $K$ increases, peaks with maximum intensity 
  develop at $\vec k=\frac23\vec b_1+\frac13\vec b_2$ and symmetry related momenta.}
  \label{StructureFactor}
\end{figure}

This scenario is confirmed by looking at the structure factor. Figure \ref{StructureFactor} shows 
$S(\vec k)$ for $18\times18$ hexagons (972 sites) with $\rho=\frac13$,  $K=0$, $K=5t$, and $K=15t$. For 
$K=0$ (left panel), only lattice Bragg peaks arise and correspond to the minima of the intensity (black and blue).
For $K=5t$ (middle panel), high intensity regions (red and yellow) appear in the center of triangles 
formed by the Bragg peaks. In the insulating phase, $K=15t$ (right panel), the high intensity regions 
become localized at $\vec k_{max}=(\frac23,\frac13)$ in the $(\vec b_1,\vec b_2)$ basis (and symmetry 
related momenta), and form a honeycomb lattice. Since $\vec k_{max}$ is parallel to $\vec a_1$ and 
$||\vec k_{max}||=\frac{4\pi}{3}$, the insulating phase has features that appear with a spatial 
frequency of $\frac23$ in the $\vec a_1$ direction, in agreement with Fig.~\ref{VBS}.

To conclude, we study hard-core bosons on a kagome lattice with a sextic ring-exchange term using 
the Stochastic Green Function algorithm~\cite{SGF,DirectedUpdate,GlobalSpaceTimeUpdate}.  
We find that the system becomes unstable towards phase separation as the ring-exchange interaction increases, 
except for densities $\rho=\frac13$ and $\rho=\frac23$ where the superfluid 
is destroyed. Here we observe an unusual valence bond solid phase with resonances involving six 
sites simultaneously. The hysteresis obtained in the quantum Monte Carlo indicates that the phase transition between 
the superfluid and the valence bond solid at $\rho=\frac13$ and $\rho=\frac23$ is first order.
Models with higher order ring-exchange played an important role in the search of spin liquid
phases due to their relation with the gauge theory deconfined phase. This work showcases the power of 
the Stochastic Green Function algorithm 
to study models with higher order ring-exchange interactions which have hitherto been very challenging 
for other Monte Carlo methods. 

\begin{acknowledgments}
We thank Carol Duran for careful reading of the manuscript.
This work is supported by DOE SciDAC grant DE-FC02-06ER25792 (KMT and MJ) and by 
NSF OISE-0952300 (VGR and JM). This work used the Extreme Science and Engineering 
Discovery Environment (XSEDE), which is supported by National Science Foundation 
grant number DMR100007.  
\end{acknowledgments}

\end{document}